\documentclass[12pt,a4paper]{article}
\usepackage{amsfonts,amsmath,amssymb,bm,graphicx}
\usepackage[english]{babel}

\begin{document}

\vskip 22pt

\begin{center}
{\bf \LARGE {
Hypermagnetic helicity evolution in early universe: leptogenesis and hypermagnetic diffusion
 }}
\end{center}

\vskip 12pt

\begin{center}
\small
{\bf V.\,B.~Semikoz$^{a}$\footnote{e-mail: semikoz@yandex.ru}, A.\,Yu.~Smirnov$^{a}$\footnote{e-mail: smirnoff.alexandr@gmail.com},~ and D.\,D.~Sokoloff$^b$\footnote{e-mail: sokoloff.dd@gmail.com} } \\
\vskip 3pt
{\em $^a${\it Pushkov Institute of Terrestrial Magnetism, Ionosphere}\\{\it and Radiowave Propagation of the Russian Academy}\\{\it of Sciences (IZMIRAN), Troitsk, Moscow, 142190 Russia}\\ } 
{\em $^b${\it Department of Physics, Moscow State University, \\{\it 119999, Moscow, Russia}} } \\
\end{center}

\vskip 20pt

\begin{abstract}
{
We study hypermagnetic helicity and lepton asymmetry evolution in plasma of the early Universe before the electroweak phase transition (EWPT) accounting for chirality flip processes via inverse Higgs decays and sphaleron transitions which violate the left lepton number and wash 
out the baryon asymmetry of the Universe (BAU). In the scenario where the right electron asymmetry supports the BAU alone through the conservation law $B/3 - L_{eR}=const$ at temperatures $T>T_{RL}\simeq 10~TeV$ the following universe cooling leads to the production of a non-zero left lepton (electrons and neutrinos) asymmetry. This is due to the Higgs decays becoming more faster when entering the equilibrium at $T=T_{RL}$ with the universe expansion, $\Gamma_{RL}\sim T> H\sim T^2$ , resulting in the parallel evolution of both the right and the left electron asymmetries at $T<T_{RL}$ through the corresponding Abelian anomalies in SM in the presence of a seed hypermagnetic field. The hypermagnetic helicity evolution proceeds in a self-consistent way with the lepton asymmetry growth. The role of sphaleron transitions decreasing the left lepton number turns out to be negligible in given scenario. The hypermagnetic helicity can be a supply for the magnetic one in Higgs phase assuming a strong seed hypermagnetic field in symmetric phase. 
\vspace{12pt}

}
\end{abstract}

\section{Introduction}

Magnetic helicity  $H=\int {\rm d}^3x{\bf A}\cdot{\bf B}$ where ${\bf B}$ is the
magnetic field and ${\bf A}$ is its vector potential is an inviscid invariant of motion in contemporary Universe. The corresponding conservation law $dH/dt=0$ severely constrains magnetic field generation by dynamo process and its further evolution (see e.g. \cite{Betal12}). Conventional models of magnetic field evolution in celestial bodies presume that an initial supply of magnetic helicity in the body was negligible and further magnetic field evolution and large-scale magnetic field formation have to redistribute somehow magnetic helicity between large-scale and small-scale fields in order to keep
total magnetic helicity vanishing. This viewpoint looks as a reasonable approach, however, the whole truth may be more complicated because the seed magnetic field can be substantial and helical. Possible influence of initial magnetic helicity supply on the further magnetic hystory of a celestial body remains practically untouched by contemporary MHD studies. 

The problem is relevant in particular for the first celestial bodies in the Universe (large-scale structure of the Universe, quasars, first galaxies). The point is that the seed magnetic fields for the first objects in the Universe may be originated in cosmological magnetic fields obtained from the Early Universe just after the Big Bang\cite{Grasso:2000wj}. A conventional viewpoint is that the Early Universe did contain a magnetic field. The option that this cosmological magnetic field was a helical one looks attractive because magnetic helicity is not an inviscid invariant of motion in the very hot cosmological plasma and its generation in such plasma may be expected. More precisely, magnetic field as well as magnetic helicity governed by standard Maxwell equations in contemporary Universe might be originated from their progenitors such as hypermagnetic field and hypermagnetic helicity evolved in the symmetric phase of electroweak plasma before the electroweak phase transition (EWPT).  In contrast to the magnetic helicity, hypermagnetic helicity is not an inviscid invariant of motion and its growth at the early stages of evolution of the Universe is not directly constrained by conservation laws.  

The first observational indications of the presence of cosmological magnetic fields in the 
inter-galactic medium which may survive even till the present epoch~\cite{Neronov:2009gh,Neronov:1900zz} were as a new incitement 
to the conception of cosmological magnetic field and helicity.  

Relying on the option of the hypermagnetic (magnetic) 
field evolution passing the EWPT time 
we reveal here the magnetic helicity generation in the early universe. 
This opens a new option that magnetic field starts from a state 
with a substantial supply of magnetic helicity. Note that our approach is based
on the $\alpha_Y$ -helicity parameter derived as a polarization effect in plasma from the Chern-Simons anomaly
in effective Lagrangian for the hypercharge field $Y_{\mu}$ \cite{Semikoz:2011tm,Giovannini:1997eg,Redlich:1984md}.
The conversion of the hypermagnetic helicity into the magnetic one during EWPT has been already considered in \cite{Akhmet'ev:2010ba}.

In the recent work \cite{Semikoz:2012ka} authors  have already studied evolution of the hypermagnetic helicity density $h_Y(t)=H_Y(t)/V$ neglecting 
hypermagnetic diffusion term that is proportional to the diffusion parameter $\eta_Y=(\sigma_{cond})^{-1}$ where $\sigma_{cond}=100~T$ is the hot plasma conductivity. Since both the helicity parameter $\alpha_Y$ and the diffusion one are inverse proportional to such conductivity the  
limit $\sigma_{cond}\to \infty$ for an ideal plasma is forbidden. Retaining $\alpha_Y$-term while neglecting the diffusion term
authors \cite{Semikoz:2012ka} relied on huge scales of hypermagnetic field surely surviving diffusion that looks
as a too rough approximation.  To avoid such approximation 
in the present work we consider the case of arbitrary scale of hypermagnetic field. 

Note that our approach resembles the analysis of magnetic helicity evolution in the Fourier representation 
applied in paper \cite{Boyarsky:2011uy} for Maxwellian magnetic fields evolved from the hypermagnetic ones at the stage after EWPT (see below in Section 2). The non-zero polarization operator in SM plasma based on different chemical potentials of left and right fermions, $\mu_L\neq \mu_R$, leads to instability in QED plus Fermi theory and corresponding generation of a large-scale Maxwellian field \cite{Boyarsky:2012ex}. We try here to determine how large such difference can be developed for leptons $\Delta \mu=\mu_{eR} - \mu_{eL}\neq 0$ before EWPT.

As a first step for the hypermagnetic helicity evolution 
we consider here the hot universe plasma before EWPT at the stage $T_{RL}>T>T_{EW}$,
when left leptons in the SM doublet $L=(\nu_{eL}e_L)^T$ enter equilibrium with right electrons
$e_R$ due to inverse Higgs decays like $e_R\bar{e}_L\to \varphi^{(0)}$, 
$e_R\bar{\nu}_{eL}\to \varphi^{(-)}$. This happens during universe
cooling just at the temperature $T_{RL}\sim 10~TeV$ when  the rate of chirality flip $\Gamma_{RL}\sim T$ 
becomes bigger than the Hubble expansion $H\sim T^2$,
$\Gamma_{RL}\geq H$. This leads to an additional polarization effect given by  left lepton macroscopic currents in a 
seed hypermagnetic field ${\bf B}_Y$, $J^{(e)}_{i5}=<\bar{\psi}_{eL}\gamma_i\gamma_5\psi_{eL}>\sim \mu_{eL}B_i^Y$,
$J_{i5}^{(\nu)}=<\bar{\nu}_{eL}\gamma_i\gamma_5\nu_{eL}>\sim \mu_{eL}B_i^Y$, where electron chemical potential 
$\mu_{eL}$ coincides with the neutrino one, $\mu_{eL}=\mu_{\nu_{eL}}$.

Accounting for the evolution of the left lepton asymmetry $(n_{eL} - n_{\bar{e}L})\sim \mu_{eL}(t)$ at temperatures $T_{EW}<T<T_{RL}$ given by the left electron chemical potential $\mu_{eL}(t)\neq 0$  which evolves due to the Abelian anomaly and undergoes sphaleron influence,
we extend also scenario \cite{Semikoz:2012ka,Giovannini:1997eg} based on the leptogenesis due to evolution of
the right electron asymmetry $(n_{eR} - n_{\bar{e}R})\sim \mu_{eR}(t)\neq 0$ alone in the same hypermagnetic fields ${\bf B}_Y\neq 0$.

The goal of the present work is a complete description
of hypermagnetic helicity density evolution $h_Y(t)=\int dkh_Y(k,t)$ down to the EWPT time
accounting for the hypermagnetic diffusion both for a monochromatic and continuous spectra of the helicity density 
$h_Y(k,t)$.

The plan of our paper is the following. In section 2 we derive the kinetic equation for the hypermagnetic density spectrum in the Fourier representation using conformal variables. Such spectrum depends on the lepton asymmetries which evolve in a self-consistent way in the same hypermagnetic field as discussed in the subsection 2.1. Then in main Section 3 we solve numerically non-linear kinetic equations for lepton asymmetries
for the simple case of the fully helical field $h_Y(k,t)=2\rho_{B_Y}(k,t)/k$ when the hypermagnetic energy density $\rho_{B_Y}(t)$  evolves itself through
$\mu_{eR}(t)$ and $\mu_{eL}(t)$. In Section 4 we consider the continuous initial spectrum of the helicity density taking into account the inverse cascade in spectra that increases the scale of hypermagnetic fields, $k^{-1}\to \infty$, while diminishes their amplitudes and helicity density itself.
Then in Section 5 we discuss our results comparing them with known results in literature. 
In Appendix A we present a more general system of evolution equations for an arbitrary helicity density spectrum obeying the known limit $h_Y(k,t)\leq 2\rho_{B_Y}(k,t)/k$
\cite{Biskamp}.

\section{Hypermagnetic helicity before EWPT}

In the rest frame of the medium as a whole the Faraday induction equation
governing hypermagnetic fields ${\bf B}_Y=\nabla\times {\bf Y}$ reads \footnote{Throughout the text we have neglected the bulk velocity evolution described by the Navier-Stokes equation since the length scale of the velocity variation $\lambda_v$ is much shorter than the correlation distance of the hypermagnetic field, $\lambda_v\ll k^{-1}$, or infrared modes of the hypermagnetic field  are practically unaffected by the velocity of plasma. In addition, the bulk velocity ${\bf v}$ does not contribute to the helicity evolution $dh_Y/dt\sim ({\bf E}_Y\cdot{\bf B}_Y)$ when the generalized Ohm law is substituted, ${\bf E}_Y= - {\bf v}\times {\bf B}_Y + \eta_Y\nabla\times {\bf B}_Y - \alpha_Y{\bf B}_Y$. }
\begin{equation}\label{Faraday}
\frac{\partial {\bf B}_Y}{\partial t}=\nabla\times \alpha_Y{\bf B}_Y + \eta_Y\nabla^2{\bf B}_Y,
\end{equation}
where at the temperatures $T_{RL}> T >T_{EW}$ the hypermagnetic helicity
coefficient $\alpha_Y$ is given by the right
and left electron chemical potentials $\mu_{eR}$, $\mu_{eL}$ \cite{Dvornikov:2011ey,Dvornikov:2012rk} \footnote{The sign for $\alpha_Y$ and $\gamma_5$ is opposite to the sign chosen in ~\cite{Giovannini:1997eg,Semikoz:2011tm} and coincides here with the definitions in \cite{Zee} where $\psi_R=(1 + \gamma_5)\psi$ is the right fermion field. See also in \cite{Dvornikov:2011ey,Dvornikov:2012rk}.},
\begin{equation}\label{alpha}
\alpha_Y(T)=+\frac{g^{'2}(\mu_{eR} + \mu_{eL}/2)}{4\pi^2\sigma_{cond}}~~,
\end{equation}
and $\eta_Y =(\sigma_{\rm cond})^{-1}$ is the hypermagnetic
diffusion coefficient, $\sigma_{\rm cond}(T) \simeq 100 T$ is the hot plasma
conductivity.  Let us stress that $\alpha_Y$ -effect in Faraday equation (\ref{Faraday})
arises due to Abelian anomalies for right and left electron (neutrino) currents both nonpersistent in
the presence of a hypercharge field $Y_{\mu}$ at the temperatures $T<T_{RL}$.  Multiplying Eq. (\ref{Faraday}) by the
corresponding vector potential and adding the analogous construction
produced by evolution equation governing the vector potential
(multiplied by hypermagnetic field) after integration over space we
get the evolution equation for the hypermagnetic helicity ${\rm
H}_Y=\int d^3x {\bf Y}\cdot{\bf B}_Y$:
\begin{eqnarray}\label{helicity}
&&\frac{{\rm dH}_Y}{{\rm dt}}=
-2\int_V({\bf E}_Y\cdot{\bf B}_Y)d^3x -\oint [Y_0{\bf B}_Y 
+{\bf E}_Y\times {\bf Y}]d^2S=\nonumber\\&&= -2\eta_Y (t)\int d^3x(\nabla\times {\bf B}_Y)\cdot {\bf B}_Y + 2\alpha_Y (t)\int d^3xB_Y^2(t).
\end{eqnarray}

For the single symmetric phase before EWPT we 
have just omitted in the last line in (\ref{helicity}) the surface integral $\oint(...)$ since
hypercharge fields vanish at infinity. However,  such surface integral can be
important at the boundaries of different phases during EWPT, $T\sim
T_{EW}$. In paper \cite{Akhmet'ev:2010ba} authors had
studied  how the hypermagnetic helicity flux penetrates  such
surface separating symmetric and broken phases and how hypermagnetic
helicity density $h_Y={\bf B}_Y{\bf Y}$ converts into the magnetic
helicity density $h={\bf B}{\bf A}$ at the EWPT time.

Note that the evolution equation (\ref{helicity}) is similar to Eq. (7) in
paper ~\cite{Semikoz:2004rr} derived in SM for times after
EWPT, $T\ll T_{EW}$. The authors used there the Fermi point-like ({\it
short-ranged}) neutrino interaction with plasma mediated by heavy
$W,Z$-bosons instead of {\it long-ranged}  interaction through the
massless hypercharge field $Y_{\mu}$ at times much above the EWPT
temperature $T_{EW}$, $T\gg T_{EW}$.  As a result the coefficents
$\eta$, $\eta_Y$ and $\alpha$, $\alpha_Y$ have different forms in
\cite{Semikoz:2004rr,Semikoz:2003qt} and in papers
~\cite{Giovannini:1997eg,Semikoz:2012ka}.

Let us change physical variables to the conformal ones using the conformal time $\eta=M_0/T$, $M_0=M_{Pl}/1.66\sqrt{g^*}$, where $M_{Pl}=1.2\times 10^{19}~GeV$ is the Plank mass, $g^*=106.75$ is the effective number of relativistic degrees of freedom. 

In FRW metric $ds^2=a^2(\eta)(d\eta^2 - d\tilde{{\bf x}}^2)$ using definitions $a=T^{-1}$, $a_0=1$ at the present temperature $T_{now}$, $d\eta=dt/a(t)$ , we input the following notations: $\tilde{k}=ka=const$ is the conformal momentum (giving a red shift for the physical one, $k\sim T=T_{now}(1 + z)$); $\xi_a(\eta)=a\mu_a =\mu_a/T$ is the dimensionless fermion asymmetry changing over time; $\tilde{{\bf B}}_Y=a^2{\bf B}_Y$, $\tilde{\bf Y}=a{\bf Y}$ are the conformal dimensionless counterparts of hypermagnetic field and hypercharge potential correspondingly.
 
It is suitable to rewrite (\ref{helicity}) using the conformal coordinate $\tilde{{\bf x}}={\bf x}/a$ for the Fourier components 
of the helicity density \footnote{Note that exponents $e^{i{\bf k}{\bf x}}=e^{i\tilde{\bf k}\tilde{\bf x}}$ coincide in Fourier integrals both in usual variables and in the conformal ones.},
$\tilde{h}_Y(\eta)\equiv \int (\tilde{\bf Y}\cdot\tilde{\bf B}_Y)d^3x/V=\int d\tilde{k} \tilde{h}_Y(\tilde{k},\eta)$, and the hypermagnetic energy density 
$\tilde{\rho}_{B_Y}(\eta)=\tilde{B}_Y^2(\eta)/2=\int d\tilde{k}\tilde{\rho}_{B_Y}(\tilde{k},\eta)$ defined as their spectra: 
\begin{eqnarray}\label{Fourier}
&&\tilde{h}_Y(\tilde{k},\eta)=\frac{\tilde{k}^2a^3}{2\pi^2 V}\tilde{{\bf Y}}(\tilde{k},\eta)\cdot\tilde{{\bf B}}_Y^*(\tilde{k},\eta),\nonumber\\&&\tilde{\rho}_{B_Y}(\tilde{k},\eta)=\frac{\tilde{k}^2a^3}{4\pi^2V}\tilde{{\bf B}}(\tilde{k},\eta)\cdot\tilde{{\bf B}}_Y^*(\tilde{k},\eta)~.
\end{eqnarray}
This allows us to calculate integrals $\int d^3x(...)/V$ in (\ref{helicity}) and results in\footnote{In standard dimensional notations $h_Y(\eta)$ and $h_Y(\tilde{k},\eta)$ are measured in units $G^2cm$ or $M^4L=M^3=L^{-3}$ that is given by the volume $V$ in (\ref{Fourier}) accounting for the relation $\tilde{h}_Y(\tilde{k},\eta)=a^3h_Y(\tilde{k},\eta)$.}
\begin{equation}\label{conformal}
\frac{d\tilde{h}_Y(\tilde{k},\eta)}{d\eta}=-\frac{2\tilde{k}^2\tilde{h}_Y(\tilde{k},\eta)}{\sigma_c} +\left(\frac{2\alpha^{'}[\xi_{eR}(\eta) + \xi_{eL}(\eta)/2]\tilde{k}}{\pi\sigma_c}\right)\tilde{h}_Y(\tilde{k},\eta),
\end{equation}
where $\alpha^{'}=g^{'2}/4\pi$ is given by the SM coupling, $\sigma_c=\sigma_{cond}a=\sigma_{cond}/T\approx 100$ is the dimensionless plasma conductivity; $\xi_{eR}(\eta)=\mu_{eR}(T)/T$ and $\xi_{eL}(\eta)=\mu_{eL}(T)/T$ are the right and left electron asymmetry correspondingly. Note that in (\ref{conformal}) we substituted the hypermagnetic energy density $\tilde{\rho}_{B_Y}(t)=\tilde{B}^2_Y(t)/2$ for the maximum helical hypermagnetic field $\tilde{\rho}_{B_Y}(\tilde{k},\eta)=\tilde{k}\tilde{h}_Y(\tilde{k},\eta)/2$ given by Eq. (\ref{Fourier}). Such choice of the fully helical hypercharge field allows to get the simple differential Eq. (\ref{conformal})
and provides an efficient inverse cascade for turbulent Maxwellian magnetic fields evolved after EWPT from hypercharge fields considered here. 

As an example of such field which is not applied here while it obeys the gauge $\nabla\cdot{\bf Y}=0$, $Y_0=0$, one can mention the Chern-Simons wave ${\bf Y}=Y(t)(\sin k_0z, cosk_0z, 0)$ for which the hypermagnetic field ${\bf B}_Y=\nabla\times {\bf Y}=k_0{\bf Y}$ has a non-trivial topology being the maximum helical field. Indeed, its helicity density $h_Y={\bf YB}_Y=k_0Y^2(t)$ is connected with the energy density $\rho_{B_Y}={\bf B}_Y^2/2=k^2_0Y^2(t)/2$ exactly through the relation $k_0h_Y/2=\rho_{B_Y}$.

The solution of Eq. (\ref{conformal}) takes the form (compare Eq. (8) in \cite{Boyarsky:2011uy}):
\begin{equation}\label{helicitysolution}
\tilde{h}_Y(\tilde{k},\eta)=\tilde{h}_Y^{(0)}(\tilde{k},\eta_0)\exp \left(\frac{2\tilde{k}}{\sigma_c}\left[\frac{\alpha^{'}}{\pi}\int_{\eta_0}^{\eta}\left(\xi_{eR}(\eta^{'}) + \frac{\xi_{eL}(\eta^{'})}{2}\right)d\eta^{'}- \tilde{k}(\eta - \eta_0)\right]\right).
\end{equation}
The spectrum of the dimensionless helicity density $\tilde{h}_Y(\tilde{k},\eta)=a^3h_Y(\tilde{k},\eta)$ can be rewritten in compact form as
 \begin{equation}\label{conformsolution}
\tilde{h}_Y(\tilde{k},\eta)\equiv\frac{h_Y(\tilde{k},\eta)}{T^3}= \tilde{h}^{(0)}_Y(\tilde{k},\eta_0)\exp \left[A(\eta)\tilde{k} -B(\eta)\tilde{k}^2\right],
 \end{equation}
 where the initial spectrum $\tilde{h}^{(0)}_Y(\tilde{k},\eta_0)=h_Y(\tilde{k},\eta_0)/T_0^3$ corresponds in our scenario to the moment of the left asymmetry appearance  at $T_0=T_{RL}$, and we used notations taken from (\ref{helicitysolution})
 \begin{equation}\label{parameter} A(\eta)=\frac{2\alpha^{'}}{\pi\sigma_c}\int_{\eta_0}^{\eta}\left(\xi_{eR}(\eta^{'}) + \frac{\xi_{eL}(\eta^{'})}{2}\right)d\eta^{'},~~~~B(\eta)=\frac{2}{\sigma_c}(\eta - \eta_0).
 \end{equation}
 
Neglecting quantum effect given by Abelian anomalies (case $\alpha^{'}=0$) and in the absence of hypermagnetic diffusion (when both dynamical effects vanish in ideal plasma limit, $\sigma_c\to \infty$) we get from (\ref{conformsolution}) the standard conservation of the helicity density $d\tilde{h}_Y/d\eta =0$, $\tilde{h}_Y=const$, with the conformal scaling $h_Y(\eta)=(\eta_0/\eta)^3h_Y(\eta_0)$.
 
 To calculate the helicity density spectrum (\ref{conformsolution}) we should find the lepton asymmetry functions $\xi_{eR}(\eta)$, $\xi_{eL}(\eta)$ developing in a self-consistent way.
 
\subsection{ Evolution of lepton asymmetries}
 For simplicity we consider inverse Higgs decays only or we neglect the Higgs boson asymmetry, $\mu_0=0$.
The system of kinetic equations for leptons accounting for Abelian anomalies for right electrons and left electrons (neutrinos), inverse Higgs decays and sphaleron transitions as well, takes the form:
\begin{eqnarray}\label{system} 
  &&\frac{{\rm d}L_{e_\mathrm{R}}}{\rm dt} = 
  \frac{g'^2}{4\pi^2s} (\mathbf{E}_\mathrm{Y} \cdot \mathbf{B}_\mathrm{Y}) + 
  2\Gamma_\mathrm{RL}
  \left\{
    L_{e_\mathrm{L}}-L_{e_\mathrm{R}}\right\},
\nonumber \\
  &&\frac{{\rm d}L_{e_\mathrm{L}}}{\rm dt} = 
  -\frac{g'^2}{16\pi^2s}(\mathbf{E}_\mathrm{Y} \cdot \mathbf{B}_\mathrm{Y}) +
  \Gamma_\mathrm{RL}
  \left\{
    L_{e_\mathrm{R}} - L_{e_\mathrm{L}}\right\} - \left(\frac{\Gamma_{sph}T}{2}\right)L_{e_\mathrm{L}}.
 \end{eqnarray}
Here $L_b=(n_b - n_{\bar{b}})/s\approx T^3\xi_b/6s$ is the lepton number, $b=e_\mathrm{R},e_\mathrm{L}, \nu_e^\mathrm{L}$, $s=2\pi^2g^*T^3/45$ is the entropy density, and $g^*=106.75$ is the number of relativistic degrees of freedom. The factor=2 in the first line takes into account the equivalent reaction branches, $e_R\bar{e}_L\to \tilde{\varphi}^{(0)}$ and $e_R\bar{\nu}_{e^\mathrm{L}}\to \varphi^{{(-)}}$; $\Gamma_{RL}$ is the chirality flip rate. Of course, for the left doublet $L_e^T=(\nu_{e^\mathrm{L}},e_L)$ kinetic equation for neutrino number is excess because $L_{e_\mathrm{L}}=L_{\nu_e^\mathrm{L}}$. Then $\Gamma_{sph}=C\alpha_W^5=C(3.2\times 10^{-8})$ is the dimensionless probability  of sphaleron transitions decreasing the left lepton numbers and therefore washing out baryon asymmetry of universe (BAU). It is given by the $SU(2)_W$ constant $\alpha_W=g^2/4\pi=1/137\sin^2\theta_W=3.17\times 10^{-2}$ where $g=e/\sin \theta_{W}$ is the gauge coupling  in SM and the constant $C\simeq 25$ is estimated through lattice calculations (see, e.g., the chapter 11 in the book \cite{Rubakov}).

In conformal variables after integration of the  system (\ref{system}) over volume $\int d^3x(...)/V$, transferring to the Fourier components for hypercharge fields the kinetic equations (\ref{system}) take the form
 \begin{equation}\label{right}
\frac{d\xi_{eR}(\eta)}{d\eta}=- \frac{3\alpha^{'}}{\pi}\int d\tilde{k} \frac{d\tilde{h}_Y(\tilde{k},\eta)}{d\eta} - \Gamma\Bigl[\xi_{eR}(\eta) - \xi_{eL}(\eta)\Bigr],
\end{equation}
\begin{equation}\label{left}
\frac{d\xi_{eL}(\eta)}{d\eta}=+ \frac{3\alpha^{'}}{4\pi}\int d\tilde{k} \frac{d\tilde{h}_Y(\tilde{k},\eta)}{d\eta} - \frac{\Gamma (\eta)}{2}\Bigl[\xi_{eL}(\eta) - \xi_{eR}(\eta)\Bigr] - \frac{\Gamma_{sph}}{2}\xi_{eL}(\eta),
\end{equation}
where 
\begin{equation}\label{rate}
\Gamma(\eta)=\left(\frac{242}{\eta_{EW}}\right)\left[1 - \left(\frac{\eta}{\eta_{EW}}\right)^2\right],~~~\eta_{RL}<\eta <\eta_{EW}
\end{equation}
is the dimensionless chirality flip rate $\Gamma=2a\Gamma_{RL}$ \cite{Dvornikov:2011ey,Campbell:1992jd} ,
$\eta_{EW}=M_0/T_{EW}=7\times 10^{15}$ is the EWPT time at $T_{EW}=100~GeV$. The derivative in the integrands of first terms in (\ref{right}), (\ref{left}), $d\tilde{h}_Y(\tilde{k},\eta)/d\eta$, is given by Eq. (\ref{conformal}) where in the right hand side we should substitute $\tilde{h}_Y(\tilde{k},\eta)$ taken from Eq. (\ref{conformsolution}).

We choose the following initial conditions at the time $\eta_0=\eta_{RL}=7\times 10^{13}$ that corresponds to the temperature $T_{RL}=10~TeV$:
\begin{equation}\label{initial}\xi_{eL}(\eta_0)=0,~~~~ \xi_{eR}(\eta_0)=10^{-10}.\end{equation}
In subsection 3.2.1 we discuss also the case of a large initial lepton asymmetry, $\xi_{eR}(\eta_0)=10^{-4}$, that is a free parameter in our problem.

The solution of the system (\ref{right}) and (\ref{left}) allows us to calculate the evolution of hypermagnetic
helicity density (\ref{conformsolution}) for the two cases considered below: a) monochromatic helicity density spectrum
\begin{equation}\label{mono}
\tilde{h}_Y(\tilde{k},\eta)=\tilde{h}_Y(\eta)\delta (\tilde{k} - \tilde{k}_0),
\end{equation} 
and b) for the continuous initial spectrum $\tilde{h}_Y(\tilde{k},\eta_0)\sim \tilde{k}^{n_s}$, $n_s\geq 3$.
Here the initial helicity density amplitude for the monochromatic spectrum (\ref{mono}) $\tilde{h}_Y(\eta_0)=(\tilde{B}^Y_0)^2/\tilde{k}_0$ is given by a seed field $\tilde{B}^Y_0$. There are the two free parameters in our problem: a) a seed hypermagnetic field $\tilde{B}^Y_0$ at the initial temperature $T_0=T_{RL}=10~TeV$ and b) a value of the initial right electron asymmetry $\xi_{eR}(\eta_0)\neq 0$ in the chosen scenario  \cite{Dvornikov:2011ey,Dvornikov:2012rk}. We choose throughout text the initial hypermagnetic energy density $\tilde{\rho}_{B_Y}^{(0)}=10^{-8}$ that corresponds to a strong seed field $B_0^Y=10^{-4}\sqrt{2}T_0^2\sim 10^{24}~G$. Note that such field does not influence the Friedman law of the Universe expansion, $\rho_{B_Y}\ll \rho_{\gamma}\sim T^4$.

\section{Lepton asymmetry and helicity evolution in monochromatic hypermagnetic field}
In this section we study evolution of lepton asymmetries for a maximum helical hypermagnetic field in the case of monochromatic helicity spectrum (\ref{mono}). 

\subsection{Saturation regime for the monochromatic spectrum (\ref{mono})}
Let us rewrite kinetic equations for asymmetries (\ref{right}), (\ref{left}) accounting for the helicity evolution (\ref{conformal}):
\begin{eqnarray}\label{leftright}
\frac{d\xi_{eL}(\eta)}{d\eta}=&&-\frac{3\alpha^{'}}{\pi \sigma_c}\int d\tilde{k}\frac{\tilde{k}^2\tilde{h}_Y(\tilde{k},\eta)}{2} + \frac{3\alpha^{'2}}{\pi^2\sigma_c}\tilde{\rho}_{B_Y}(\eta)\left(\xi_{eR} + \frac{\xi_{eL}}{2}\right) \nonumber\\&&- \frac{\Gamma}{2}(\xi_{eL} - \xi_{eR})- \frac{\Gamma_{sph}}{2}\xi_{eL}(\eta),\nonumber\\
\frac{d\xi_{eR}(\eta)}{d\eta}=&&\frac{12\alpha^{'}}{\pi \sigma_c}\int d\tilde{k}\frac{\tilde{k}^2\tilde{h}_Y(\tilde{k},\eta)}{2} -\frac{12\alpha^{'2}}{\pi^2\sigma_c}\tilde{\rho}_{B_Y}(\eta)\left(\xi_{eR} + \frac{\xi_{eL}}{2}\right) \nonumber\\&&-\Gamma(\xi_{eR} - \xi_{eL}).\nonumber\\
\end{eqnarray}
Here in the second term ($\sim \tilde{\rho}_{B_Y}=\tilde{B}^2/2=B^2a^4/2$) we used the relation for the maximum helical field $\tilde{k}\tilde{h}(\tilde{k},\eta)=2\tilde{\rho}_{B_Y}(\tilde{k},\eta)$ and definition of the total dimensionless  energy density $\tilde{\rho}_{B_Y}(\eta)=\int d\tilde{k}\tilde{\rho}_{B_Y}(\tilde{k},\eta)$. For the monochromatic field and its helicity (\ref{mono}),
for which the relation $\tilde{\rho}_{B_Y}(\tilde{k},\eta)=\tilde{k}\tilde{h}_Y(\eta)\delta (\tilde{k} - \tilde{k}_0)/2=\tilde{\rho}_{B_Y}(\eta)\delta (\tilde{k} - \tilde{k}_0)$ allows to calculate integral in first terms in the r.h.s. (\ref{leftright}), we obtain simple differential equations

\begin{eqnarray}\label{leftright2}
\frac{d\xi_{eL}(\eta)}{d\eta}=&&\left[-\frac{3\alpha^{'}\tilde{k}_0}{\pi \sigma_c} + \frac{3\alpha^{'2}}{\pi^2\sigma_c}\left(\xi_{eR} + \frac{\xi_{eL}}{2}\right)\right]\tilde{\rho}_{B_Y}(\eta)  \nonumber\\
&&-\frac{\Gamma}{2}(\xi_{eL} - \xi_{eR})- \frac{\Gamma_{sph}}{2}\xi_{eL}(\eta),\nonumber\\
\frac{d\xi_{eR}(\eta)}{d\eta}=&&\left[\frac{12\alpha^{'}\tilde{k}_0}{\pi \sigma_c} -\frac{12\alpha^{'2}}{\pi^2\sigma_c}\left(\xi_{eR} + \frac{\xi_{eL}}{2}\right)\right]\tilde{\rho}_{B_Y}(\eta) - \Gamma(\xi_{eR} - \xi_{eL}).\nonumber\\
\end{eqnarray}
Multiplying the first equation by the factor 4, then adding both equations one gets for the saturation regime $\partial_{\eta}\xi_{eR}=\partial_{\eta}\xi_{eL}\approx 0$ 
the asymmetry relation 
\begin{equation}\label{saturation2}
\xi_{eL}=\frac{\Gamma\xi_{eR}}{\Gamma + 2 \Gamma_{sph}}\ll \xi_{eR},
\end{equation}
where $\Gamma_{sph}\gg \Gamma$.
\subsubsection{Evolution to saturation for monochromatic helicity density}
Let us rewrite kinetic equations for $\xi_{eR}(\eta)$, $\xi_{eL}(\eta)$ (\ref{leftright2}) as coupled
evolution equations for the difference $\Delta \xi_e=\xi_{eR} - \xi_{eL}$ and the sum $\Xi_e=\xi_{eR} + \xi_{eL}/2$:
\begin{eqnarray}\label{difsum}
\frac{d\Delta \xi_e (\eta)}{d\eta}=&&- \frac{15\alpha^{'2}\tilde{\rho}_{B_Y}(\eta)}{\pi^2\sigma_c}\left[\Xi_e(\eta) - \Xi_e^{(satur)}\right] \nonumber\\&&-\left[\frac{3\Gamma (\eta)}{2} + \frac{\Gamma_{sph}}{3}\right]\Delta \xi_e(\eta) + \frac{\Gamma_{sph}}{3}\Xi_e(\eta),            \nonumber\\                
\frac{d\Xi_e(\eta)}{d\eta}=&& -\frac{21\alpha^{'2}\tilde{\rho}_{B_Y}(\eta)}{2\pi^2\sigma_c}\left[\Xi_e(\eta) - \Xi_e^{(satur)}\right] \nonumber\\&&-\left[\frac{3}{4}\Gamma (\eta) - \frac{\Gamma_{sph}}{6}\right]\Delta \xi_e(\eta) - \frac{\Gamma_{sph}}{6}\Xi_e(\eta),
\end{eqnarray}
where for the single  mode (\ref{mono}) the saturation sum of asymmetries $\Xi_e^{(satur)}$ is given by the constant value coming from (\ref{helicitysolution}) when exponential amplification ceases, $e^0=1$:
\begin{equation}\label{saturation}
\Xi_e^{(satur)}=\frac{4\pi^2\tilde{k}_0}{g^{'2}}.
\end{equation}
The system (\ref{difsum}) is completed by the hypermagnetic energy evolution
\begin{equation}\label{energyevolution}
\frac{d\tilde{\rho}_{B_Y}(\eta)}{d\eta}=\frac{\tilde{\rho}_{B_Y}(\eta)}{\eta_{\sigma}}\left[\frac{\Xi_e(\eta)}{\Xi_e^{(satur)}} - 1\right],
\end{equation}
derived from the relation for the maximum helical field and substituting Eq. (\ref{conformal}) into
$$
\frac{d\tilde{\rho}_{B_Y}(\eta)}{d\eta}=\frac{1}{2}\int\tilde{k}d\tilde{k}\delta (\tilde{k} - \tilde{k}_0)\frac{d\tilde{h}_Y(\eta)}{d\eta}.
$$
Here $\eta_{\sigma}=\sigma_c/2\tilde{k}_0^2$ is the conformal diffusion time. 

The simple solution of Eq. (\ref{energyevolution})
\begin{equation}\label{solutionenergy}
\tilde{\rho}_{B_Y}(\eta)=\tilde{\rho}_{B_Y}^{(0)}\exp \left(\frac{1}{\eta_{\sigma}}\int_{\eta_0}^{\eta}\left[\frac{\Xi_e(\eta^{'})}{\Xi_e^{(satur)}} -1\right]d\eta^{'}\right)
\end{equation}
is consistent with what follows from the dynamo approach (see just below).
\subsubsection{Relation with dynamo for monochromatic spectrum}
Note that the energy density $\tilde{\rho}_{B_Y}(\eta)=\tilde{B}_Y^2(\eta)/2$ given by the solution (\ref{helicitysolution}) for helicity,
\begin{eqnarray}\label{energydensity2}
&&\tilde{\rho}_{B_Y}(\eta)=\frac{1}{2}\int \tilde{k}d\tilde{k}\tilde{h}_Y(\eta)\delta(\tilde{k} - \tilde{k}_0)=\frac{\tilde{h}_Y^{(0)}\tilde{k_0}}{2}\times\nonumber\\&&\times\exp \left(\frac{2\tilde{k}_0}{\sigma_c}\left[\frac{\alpha^{'}}{\pi}\int_{\eta_0}^{\eta}\Xi_e(\eta^{'})d\eta^{'} -\tilde{k}_0(\eta - \eta_0)\right]\right)=\nonumber\\&&=\frac{\tilde{h}_Y^{(0)}\tilde{k_0}}{2}\exp \left(\frac{1}{\eta_{\sigma}}\int_{\eta_0}^{\eta}\left[\frac{\Xi_e(\eta^{'})}{\Xi_e^{(satur)}} -1\right]d\eta^{'}\right)
,
\end{eqnarray}
is consistent with our old dynamo formula for $B_Y(t)$ \cite{Dvornikov:2011ey,Semikoz:2009ye,1983flma....3.....Z,Semikoz:2007ti} given by
\begin{eqnarray}\label{dynamo}
\tilde{B}_Y(k_0,t)&=&\tilde{B}_0^Y\exp\left(\int_{t_0}^t\left[\alpha_Y(t^{'})k_0a - \frac{k^2_0a^2}{\sigma_{cond}a}\right]\frac{dt^{'}}{a}\right)\nonumber\\&=& \tilde{B}_0^Y\exp\left(\frac{\tilde{k}_0}{\sigma_c}\left[\frac{\alpha^{'}}{\pi}\int_{\eta_0}^{\eta}\Xi_e(\eta^{'})d\eta^{'}-\tilde{k}_0(\eta - \eta_0)\right]\right)
\nonumber\\&=&
\tilde{B}_0^Y\exp \left(\frac{1}{2\eta_{\sigma}}\int_{\eta_0}^{\eta}\left[\frac{\Xi_e(\eta^{'})}{\Xi_e^{(satur)}} -1\right]d\eta^{'}\right),
\end{eqnarray}
where $\tilde{B}(\eta)=a^2B(\eta)$, $\eta_Y=(\sigma_{cond})^{-1}$, and relations $k_0a=\tilde{k}_0$, $d\eta^{'}=dt^{'}/a$, $\sigma_{cond}a=\sigma_c=100$ were substituted. Here relations \begin{equation}\label{initialenergy}(\tilde{B}_0^Y)^2=\tilde{k}_0\tilde{h}_Y^{(0)}=\tilde{k}_0\tilde{Y}_0\tilde{B}_0^Y=2\tilde{\rho}_{B_Y}^{(0)}\end{equation} correspond to the maximum  
helical field at the initial time $\eta_0$. 
\begin{figure}
  \centering
  \includegraphics[scale=0.6]{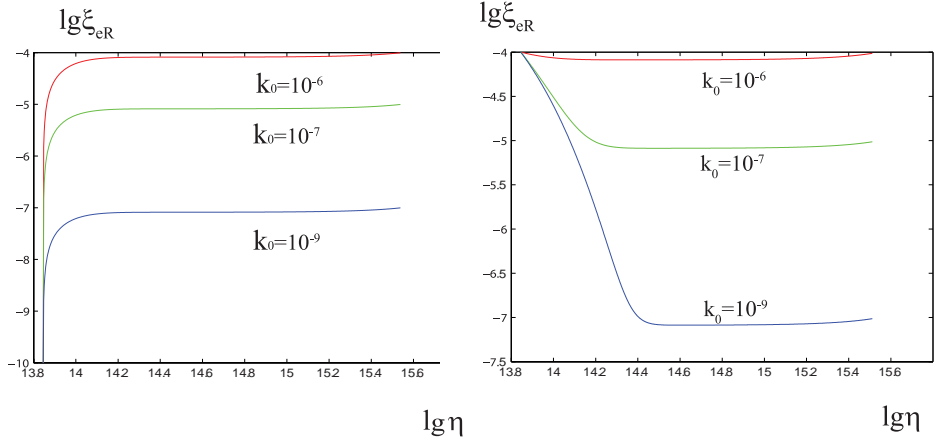}
  \caption{The evolution of the right electron asymmetry $\xi_{eR}(\eta)$ for the monohromatic spectrum (\ref{mono}).
In the left panel: the asymmetry growth for the initial value $\xi_{eR}(\eta_0)=10^{-10}$; in the right panel: the asymmetry decay for the big initial value $\xi_{eR}(\eta_0)=10^{-4}$. For the wave number $\tilde{k}_0=10^{-6}$ it is shown by the red line, for $\tilde{k}_0=10^{-7}$ by the green line and for  $\tilde{k}_0=10^{-9}$       by the blue line. }
  \label{asymmetryright}
  \end{figure}

\subsection{Solution of kinetic equations for monochromatic spectrum  (\ref{mono})} 
\begin{figure}[hb]
  \centering
  \includegraphics[scale=0.6]{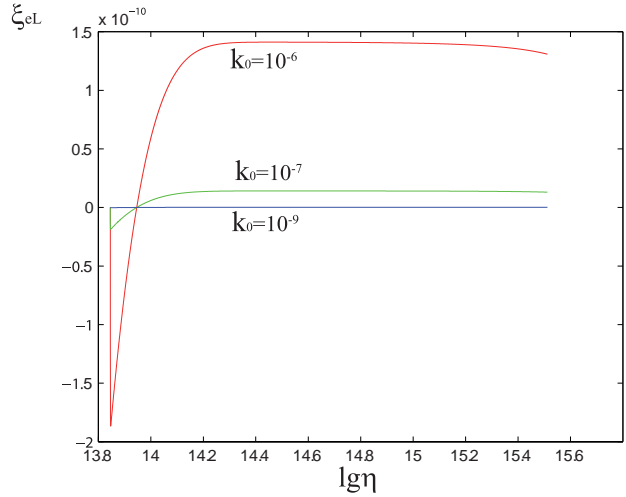}
  \caption{The evolution of the left electron asymmetry $\xi_{eL}(\eta)$ for the monohromatic spectrum (\ref{mono}).
   For the wave number $\tilde{k}_0=10^{-6}$ it is shown by the red line, for $\tilde{k}_0=10^{-7}$ by the green line and for  $\tilde{k}_0=10^{-9}$       by the blue line.}
  \label{asymmetryleft}
\end{figure}

In the left panel in Fig. \ref{asymmetryright}, and in Fig. \ref{asymmetryleft} we show solutions of kinetic equations (\ref{right}),(\ref{left}) for the right electron asymmetry $\xi_{eR}(\eta)$ 
and the left one, $\xi_{eL}(\eta)$, obeying the initial conditions (\ref{initial}) and evolving in hypermagnetic fields which are fully helical.
One can see in the left panel in Fig. \ref{asymmetryright} a sharp growth of the right electron asymmetry due to the Abelian anomaly  starting from the initial value $\xi_{eR}(\eta_0)=10^{-10}$ chosen at the level of baryon asymmetry at $T_0=T_{RL}=10~TeV$(compare in \cite{Dvornikov:2012rk}). 

On the other hand, one can choose a bigger value of the initial lepton asymmetry, let us say, $\xi_{eR}(\eta_0)\simeq 10^{-4}$.  In particular, from the considerations in $\nu$MSM -model \cite{Shaposhnikov:2008pf} one expects that the lepton asymmetry prevails over the baryon one, $\Delta L/\Delta B\geq 3\times 10^5$, or it occurs at the level $\sim 10^{-4}$. Although the required in \cite{Shaposhnikov:2008pf} lepton asymmetry must exist at temperatures $O(1)~GeV$, or in the Higgs phase after EWPT, we may assume such lepton asymmetry existing in symmetric phase too. In this case we can find oppositely to profiles shown in the left panels in Fig.\ref{asymmetryright} and Fig.\ref{helicitymono} a decrease of the lepton asymmetry shown in the right panel in Fig.\ref{asymmetryright} in parallel with the initial {\it growth} of the hypermagnetic helicity, $d\tilde{h}_Y/d\eta>0$, ended by its decay due to hypermagnetic diffusion, $d\tilde{h}_Y/d\eta<0$, shown in the right panel in Fig.\ref{helicitymono} for the mode $\tilde{k}_0=10^{-8}$. 

One can see comparing left and  right panels in Fig. \ref{asymmetryright} that the saturation level $\xi_{eR}\approx \Xi_e^{(satur)}$ remains the same one for the same wave numbers. Such independence of $\xi_{eR}(\eta)$ for $\eta\gg \eta_0$ on an initial $\xi_{eR}(\eta_0)$ stems from the analytic solution of the linear differential equation in the second line (\ref{leftright2}):
\begin{equation}\label{linear}
\frac{{\rm d}\xi_{eR}}{{\rm d}\eta}+ \left(\Gamma + \Gamma_{B_Y}\right)\xi_{eR}=Q,
\end{equation}
where we neglected $\xi_{eL}(\eta)\approx 0$ decoupling both equations (\ref{leftright2}) since $\xi_{eL}\ll \xi_{eR}$, put $Q=6\alpha^{'}\tilde{k}_0\tilde{B}_Y^2(\eta)/\pi\sigma_c\approx constant$, $\Gamma_{B_Y}=6\alpha^{'2}\tilde{B}_Y^2(\eta)/\pi^2\sigma_c\approx C_2$ when $\tilde{B}_Y(\eta)\approx C_3$ because hypermagnetic field is almost frozen in plasma, $B_Y=C_3T^2$. Then for $\eta\gg \eta_0$ when $\Gamma_{B_Y}\eta_{EW}\gg 1$ we find from (\ref{linear}):
\begin{eqnarray}\label{asymptotics}
&&\xi_{eR}(\eta)=\xi_{eR}(\eta_0)e^{-(\Gamma + \Gamma_{B_Y})(\eta - \eta_0)} + \frac{Q}{\Gamma + \Gamma_{B_Y}}\left[1 - e^{-(\Gamma + \Gamma_{B_Y})(\eta - \eta_0)}\right]\approx\nonumber\\&&\approx \frac{Q}{\Gamma + \Gamma_{B_Y}}\simeq \frac{Q}{\Gamma_{B_Y}}=\Xi_e^{(satur)},
\end{eqnarray}
where at the last step in (\ref{asymptotics}) we consider the case of strong hypermagnetic fields, $\Gamma_{B_Y}\gg \Gamma$, for which the ratio $Q/\Gamma_{B_Y}$ does not depend on a value $B_Y$ and equals to the lepton asymmetry saturation in Eq. (\ref{saturation}), $\Xi_e^{(satur)}=\pi\tilde{k}_0/\alpha^{'}$, seen in both panels. This allows us to identify similar asymptotics of $\xi_{eR}(\eta)$ for the different initial conditions $\xi_{eR}(\eta_0)=10^{-10}$ and $\xi_{eR}(\eta_0)=10^{-4}$.

In Fig. \ref{asymmetryleft} one finds a negligible growth of the left lepton asymmetry $\xi_{eL}(\eta)$ (starting from $\xi_{eL}(\eta_0)=0$) which remains at a small level near the EWPT time, $\xi_{eL}\ll \xi_{eR}$, in accordance with the estimate (\ref{saturation2}) due to the asymmetry saturation, $\partial_t\xi_{eL,eR}\approx 0$. Note that from the beginning at $T\sim T_0$ near $\xi_{eL}\approx 0$ sphaleron transitions are not efficient. As a result $\xi_{eL}(\eta)$ becomes even negative due to the negative sign of the Abelian anomaly contribution in the second kinetic Eq. (\ref{system}). One can see that the corresponding negative diffusion term within brackets in the first equation in (\ref{leftright2}) is the main one, especially,  for larger wave number $\sim \tilde{k}_0= 10^{-6}$, see negative spike in Fig. \ref{asymmetryleft}. Then for a negative $\xi_{eL}<0$ the positive term $\sim -\Gamma_{sph}\xi_{eL}$ in (\ref{leftright2}) sharply changes the sign of the derivative, $d\xi_{eL}/d\eta >0$, and the left lepton asymmetry becomes positive itself, $\xi_{eL}>0$. Following regime of saturation for $\xi_{eL}>0$ due to $\sim d\tilde{h}_Y/d\eta\approx 0$ (within brackets in (\ref{leftright2})) is slightly violated by a negative sphaleron influence the left fermion number  near the EWPT time $\eta\sim \eta_{EW}$,$d\xi_{eL}/d\eta <0$.

\begin{figure}
  \centering
  \includegraphics[scale=.6]{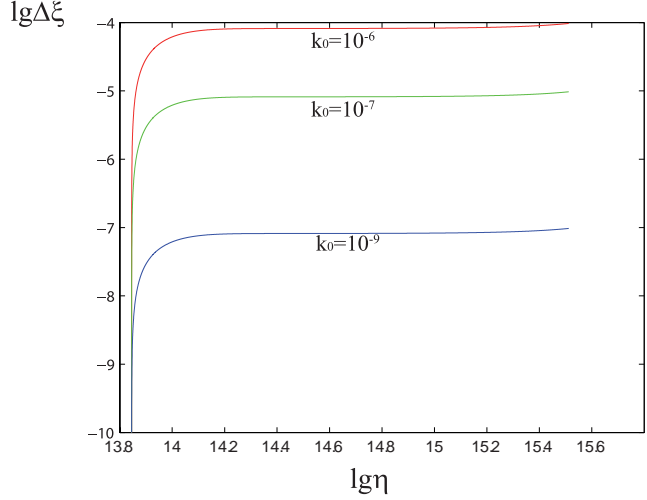}
  \caption{The chiral asymmetry parameter $\Delta \xi_e(\eta)=\xi_{eR} - \xi_{eL}$ in logarithmic scale for monochromatic helicity spectrum (\ref{mono}). For the wave number $\tilde{k}_0=10^{-6}$ it is shown by the red line, for $\tilde{k}_0=10^{-7}$ by the green line and for  $\tilde{k}_0=10^{-9}$ by the blue line.}
  \label{deltaxi}
\end{figure}

\begin{figure}
  \centering
  \includegraphics[scale=.6]{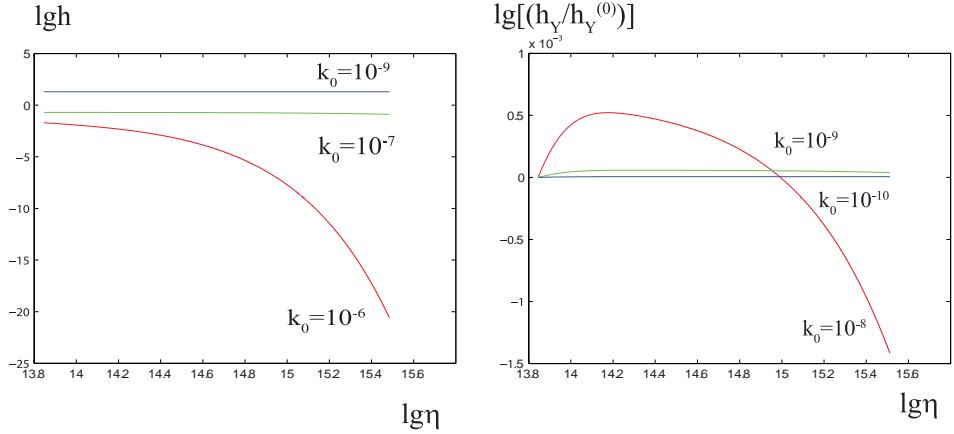}
  \caption{The helicity density modes $\tilde{h}_Y(\eta)$ for the monochromatic helicity spectrum (\ref{mono}). In the left panel: $\tilde{h}_Y(\eta)$ for the initial asymmetry $\xi_{eR}(\eta_0)=10^{-10}$. For the wave number $\tilde{k}_0=10^{-6}$ evolution is shown by the red line, for $\tilde{k}_0=10^{-7}$ by the green line and for  $\tilde{k}_0=10^{-9}$ by the blue line. In the right panel: the evolution of the normalized helicity density $\tilde{h}_Y(\eta)/\tilde{h}(\eta_0)$ for the monochromatic spectrum (\ref{mono}) and the big initial lepton asymmetry value $\xi_{eR}(\eta_0)=10^{-4}$. For the wave number $\tilde{k}_0=10^{-8}$ it is shown by the red line, for $\tilde{k}_0=10^{-9}$ by the green line and for  $\tilde{k}_0=10^{-10}$ by the blue line.}
  \label{helicitymono}
\end{figure}

 In Fig. \ref{deltaxi} we plot the evolution of the difference $\Delta \xi_e(\eta)=\xi_{eR}(\eta) - \xi_{eL}(\eta)$ which is the important initial parameter for a chiral magnetic effect after EWPT, $\Delta \mu (\eta_{EW})/T\equiv \Delta \xi_e(\eta_{EW})$ \cite{Boyarsky:2011uy,Tashiro:2012mf}.
We confirm, e.g., the growth of the chiral anomaly parameter $y_R - y_L=10^4\Delta \xi_e$ found in paper \cite{Dvornikov:2012rk} for the particular case of the Chern-Simons wave up to the value of the order $0.1\div 1.0$ for large wave numbers $\tilde{k}_0=10^{-7}\div 10^{-6}$.
 
 \subsubsection{Hypermagnetic helicity evolution for modes $\tilde{k}_0=const$}
We find from Eq. (\ref{conformsolution}) the evolution of the dimensionless hypermagnetic helicity density  $h\equiv\tilde{h}_Y(\tilde{k}_0, \eta)=h_Y(\tilde{k}_0,\eta)/T^3$   plotted in Fig.\ref{helicitymono} . One can see a decrease of the helicity density mode for the largest wave number $\tilde{k}_0=10^{-6}$. Furthermore, the longer wave length $\tilde{k}_0^{-1}$ is considered the less diffusion influences: helicity density is almost constant. In the left panel in Fig.\ref{helicitymono} we plotted along y-axis ${\rm lg\tilde{h}_Y(\eta)}={\rm lg\tilde{h}_Y(\eta_0)} + A(\eta)\tilde{k}_0 - B(\eta)\tilde{k}_0^2$, where $A(\eta)$, $B(\eta)$ are given by Eq. (\ref{parameter}). The initial helicity value seen in the left panel in Fig.\ref{helicitymono},~ ${\rm lg\tilde{h}_Y(\eta_0)}={\rm lg[(\tilde{B}^Y_0)^2/\tilde{k}_0]}$, is given by a fully helical hypermagnetic field with its initial value squared $(\tilde{B}_0^Y)^2=2\times 10^{-8}$ fixed throughout text. The different initial values $\tilde{h}_Y(\eta_0)=(\tilde{B}_0^Y)^2/\tilde{k}_0$ are bigger
for a smaller mode $\tilde{k}_0$ if the seed field $\tilde{B}_0^Y=const$ is fixed.  In the right panel in Fig.\ref{helicitymono} we plotted other curves having a different dependence of the normalized helicity $\tilde{h}_Y(\eta)/\tilde{h}_Y(\eta_0)$ on time $\eta$ calculated  for $\xi_{eR}(\eta_0)=10^{-4}$. For $\tilde{k}_0=10^{-7}$ the initial growth of helicity density is too big to be placed in the right panel while the following decay even sharper than shown here for $\tilde{k}_0=10^{-8}$.

For the case $\xi_{eR}(\eta_0)=10^{-4}$ the profiles of our chiral parameter $\Delta \xi_e$ and the helicity density $h_Y$ are similar to the analogous $\Delta \mu/T$ and $H_k$ curves for Maxwellian magnetic fields in \cite{Boyarsky:2011uy}.

Notice that the sign of the helicity density (\ref{helicitysolution}) is positive since the initial value $\tilde{h}_Y(\eta_0)>0$ when $\tilde{k_0}>0$.
The latter sign, $\tilde{k}_0>0$, is chosen to avoid the decay of the magnetic field (\ref{dynamo}) and its energy at any $\mid\tilde{k}_0\mid$ (compare after Eq. (9) in \cite{Joyce:1997uy}).

\section{Continuous helicity spectrum}
Let us denote the helicity density $\tilde{h}_Y(\eta)=\int d\tilde{k}\tilde{h}(\tilde{k},\eta)$ as the dimensionless integral over the whole Fourier spectrum $\tilde{h}_Y(\eta)=\int_0^{\tilde{k}_{max}}d\tilde{k}\tilde{h}(\tilde{k},\eta)$. Here we have extrapolated the lower limit which for $\tilde{k}\to 0$ violates the causal lower limit, $\tilde{k}>\tilde{k}_{min} =\tilde{l}_H^{-1}$, where $l_H$ is the horizon size and $\tilde{l}_H^{-1}=10^{-16}$ at $T=T_{EW}$.
The helicity density $\tilde{h}_Y(\eta)$
can be found from the evolution of helicity density in Eq. (\ref{helicitysolution}):
\begin{equation}\label{Phi}
\tilde{h}_Y(\eta)=\int_0^{\tilde{k}_{max}}\tilde{h}_Y^{(0)}(\tilde{k},\eta_0)\exp\left[\frac{1}{\eta_{\sigma}\Xi^{satur}(\tilde{k})}\int_{\eta_0}^{\eta}\left(\Xi_e(\eta^{'}) - \Xi^{sat}(\tilde{k})\right)d\eta^{'}\right]d\tilde{k}.
\end{equation}
The function $\Xi^{sat}(\tilde{k})=\pi\tilde{k}/\alpha^{'}\equiv 4\pi^2\tilde{k}/g^{'2}$ corresponds to the saturation level of the combined lepton asymmetry $\Xi_e=\xi_{eR} + \xi_{eL}/2$ for the current wave number $\tilde{k}$ . 
For the continuous spectrum \begin{equation}\label{initial2}\tilde{h}_Y(\tilde{k},\eta_0)=C\tilde{k}^{n_s}\end{equation} we have calculated the helicity density (\ref{Phi}) as
\begin{equation}\label{Phi2}
\tilde{h}_Y(\eta)=C\int _{0}^{\tilde{k}_{max}}\tilde{k}^{n_{s} } \exp \left[A(\eta)\tilde{k}-B(\eta)\tilde{k}^{2} \right]d\tilde{k}=CI_{n_s}(\eta). 
\end{equation} 
Here the functions $A(\eta)$, $B(\eta)$ are given by Eq. (\ref{parameter}).
The constant $C$ can be estimated using the relation for the fully helical field, $\tilde{h}_Y(\tilde{k},\eta_0)=C\tilde{k}^{n_s}=2\tilde{\rho}_{B_Y}(\tilde{k},\eta_0)/\tilde{k}$. Using definition of the initial
hypermagnetic energy $\int d\tilde{k}\tilde{\rho}_{B_Y}(\tilde{k},\eta_0)=(\tilde{B}_{0}^Y)^2/2$ one obtains the relation
$$C\int_0^{\tilde{k}_{max}}\tilde{k}^{n_s + 1}d \tilde{k}=(\tilde{B}_0^Y)^2=2\tilde{\rho}^{(0)}_Y=2\times 10^{-8}$$ for a finite seed field chosen above. Then we vary the maximum mode $\tilde{k}_{max}$ given by hypermagnetic diffusion. Thus, we define the constant above $C=(n_s + 2)(\tilde{B}_0^Y)^2/(\tilde{k}_{max})^{n_s +2}$.
 
\begin{figure}
  \centering
  \includegraphics[scale=.6]{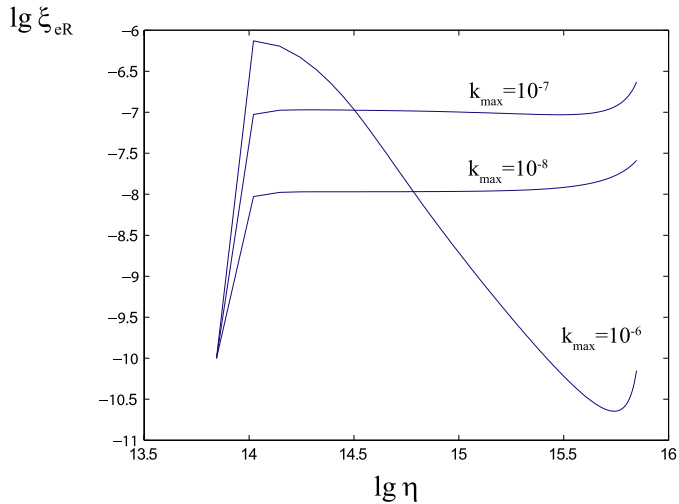}
  \caption{The evolution of the right electron asymmetry $\xi_{eR}(\eta)$ for the continuous spectrum (\ref{initial2}) and $n_s=3$. All three curves for the different upper limits in (\ref{Phi2}): $\tilde{k}_{max}=10^{-6},~10^{-7},~10^{-8}$ start from the initial value $\xi_{eR}(\eta_0)=10^{-10}$.}
  \label{asymmetryright3}
\end{figure}

For the continuous initial spectrum (\ref{initial2}) we can rewrite the kinetic equations for lepton asymmetries (\ref{right}), (\ref{left}) as
\begin{equation}\label{right2}
\frac{d\xi_{eR}}{d\eta}=\frac{6\alpha^{'}C}{\pi\sigma_c}\left[I_{n_s +2}(\eta) - \frac{\alpha^{'}}{\pi}\left(\xi_{eR} + \frac{\xi_{eL}}{2}\right)I_{n_s +1}(\eta)\right] - \Gamma(\eta)(\xi_{eR} - \xi_{eL}),
\end{equation}
\begin{eqnarray}\label{left2}
\frac{d\xi_{eL}}{d\eta}=&&-\frac{3\alpha^{'}C}{2\pi\sigma_c}\left[I_{n_s +2}(\eta) - \frac{\alpha^{'}}{\pi}\left(\xi_{eR} + \frac{\xi_{eL}}{2}\right)I_{n_s +1}(\eta)\right] - \nonumber\\&&-\Gamma(\eta)(\xi_{eL} - \xi_{eR}) - \frac{\Gamma_{sph}}{2}\xi_{eL}(\eta).
\end{eqnarray}
The integrals $I_{(n_s +2),(n_s +1)}(\eta)$ are the functions of lepton asymmetries $\xi_{eR},\xi_{eL}$ through $A(\eta)$ in Eq. (\ref{Phi2}), therefore these differential equations are strongly non-linear. 

In Fig. \ref{asymmetryright3} we show the evolution of the right electron asymmetry $\xi_{eR}(\eta)$ found from the solution of the system (\ref{right2}), (\ref{left2}) for the initial asymmetries $\xi_{eR}(\eta_0)=10^{-10}$, $\xi_{eL}(\eta_0)=0$ using the continuous helicity spectrum (\ref{initial2}) for $n_s=3$. From the beginning $\xi_{eR}$ grows for any upper limit in the integral (\ref{Phi2}). Comparing curves in that Fig.\ref{asymmetryright3} with the corresponding ones in Fig. \ref{asymmetryright} one finds a decrease of asymmetries (on $\approx$ two orders of magnitude) due to the inverse cascade of modes, $\tilde{k}_{max}\geq \tilde{k}\rightarrow 0$, in the case of the continuous spectrum Eq. (\ref{initial2}) instead of the monochromatic  one in Eq. (\ref{mono}). The same inverse 
cascade provided by integration in (\ref{Phi2}) instead of $\delta$-function entering Eq. (\ref{mono}) leads to different profiles of the curve for
$\tilde{k}_0=10^{-6}$ in Fig. \ref{asymmetryright} and the curve $\tilde{k}_{max}=10^{-6}$ in Fig. \ref{asymmetryright3} having a sharp slope somewhere at $\eta > 10^{14}$. This is due to an accumulation of the hypermagnetic diffusion effect that diminishes a growth of $\xi_{eR}$ when we integrate in Eq. (\ref{Phi2}) over wave numbers, $0<\tilde{k}<\tilde{k}_{max}$, contrary to the use of the $\delta$ - function, $\delta (\tilde{k} - \tilde{k}_0)$, for monochromatic spectrum. In such case Higgs decays lead to a sharp decrease of $\xi_{eR}$ while just before EWPT at $\eta\approx \eta_{EW}$ their contribution vanishes as $\Gamma (\eta)\to 0$ in Eq. (\ref{rate}). This leads to some growing tails seen in Fig.\ref{asymmetryright3}. Note that the hypermagnetic diffusion effect is efficient namely for a large $\tilde{k}_{max}=10^{-6}$ there since the diffusion time $\eta_{\sigma}(\tilde{k}_{max})=\sigma_c/2\tilde{k}^2_{max}=5\times 10^{13}$ is even less than the initial conformal time in our scenario,$\eta_{\sigma}(\tilde{k}_{max})<\eta_{0}=7\times 10^{13}$, when Higgs (inverse) decays enter the equilibrium with the Hubble expansion, $\Gamma\sim H$. However, for larger scales of hypermagnetic field, e.g. for $0<\tilde{k}<\tilde{k}_{max}=10^{-8}$, such diffusion time is beyond the full time interval we consider here, $\eta_{\sigma}=\sigma_c/2\tilde{k}^2\geq\sigma_c/2\tilde{k}^2_{max}=5\times 10^{17}>\eta_{EW}=7\times 10^{15}$. Therefore the corresponding profiles of $\xi_{eR}(\eta)$ for larger scales $\Lambda=\tilde{k}^{-1}$ are almost flat. 

\begin{figure}
  \centering
  \includegraphics[scale=.6]{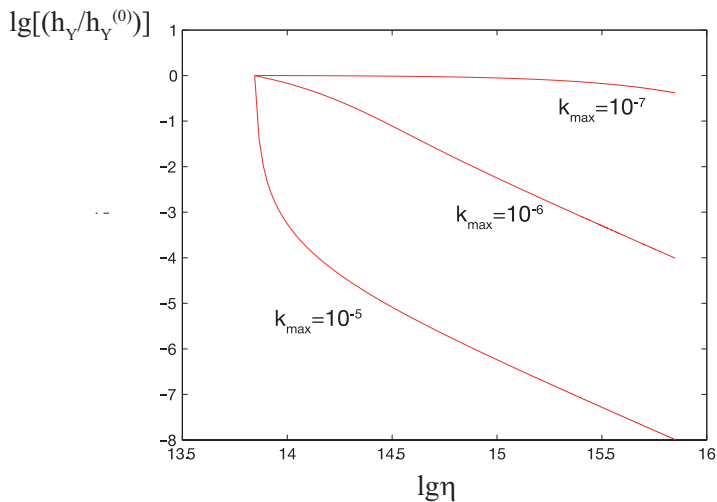}
  \caption{The evolution of the normalized hypermagnetic helicity $\tilde{h}(\eta)/\tilde{h}(\eta_0)$ for the continuous spectrum (\ref{initial2}) and $n_s=3$. All three curves for the different upper limits $\tilde{k}_{max}$ in (\ref{Phi2}) are plotted using solutions of the non-linear differential Eqs. (\ref{right2}), (\ref{left2}) with the initial lepton asymmetry values $\xi_{eR}(\eta_0)=10^{-10}$, $\xi_{eL}(\eta_0)=0$.}
  \label{helicity_contin}
\end{figure}

In Fig. \ref{helicity_contin} we show the evolution of the normalized helicity density (\ref{Phi2}) for the continuous spectrum $n_s=3$.  The exhaustion of the initial helicity density $\tilde{h}_Y(\eta_0)=(n_s+2)(\tilde{B}_0^Y)^2/[(n_s + 1)\tilde{k}_{max}]$   due to hypermagnetic diffusion before EWPT seen in given plot for large $\tilde{k}_{max}$ depends on a choice of $\tilde{k}_{max}$. For large hypermagnetic field scales $\Lambda=k^{-1}$, or for $\tilde{k}^{-1}\geq (\tilde{k}_{max})^{-1}=10^7$  one can expect the conservation of an initial helicity density, $\tilde{h}_Y(\eta)\approx \tilde{h}_Y(\eta_0)$.  We estimate the corresponding critical helicity value as $\tilde{h}_Y(\eta_0)=1/16$ for $n_s=3$ substituting the fixed initial energy density $\tilde{\rho}_{B_Y}(\eta_0)= 10^{-8}$. In dimensional units this corresponds to $h_Y(\eta_0)=3\times 10^{36}~{\rm G^2cm}$ for  $h_Y=\tilde{h}_YT^3$ at $T=T_0$ . Such value is much bigger than the galactic magnetic helicity density $h_{gal}\simeq 10^{11}~{\rm G^2cm}$. However, the following magnetic helicity conservation $d\tilde{h}/d\eta=0$ leads to a strong decrease $h=h_{EW}(\eta_{EW}/\eta)^3$ at $\eta\gg \eta_{EW}$. Thus, it could to be impossible to provide a seed galactic magnetic helicity without additional supply after EWPT, e.g. due to a matter motion (velocity fluxes) we neglected here. Such approach is beyond the scope of the present work.

\section{Discussion}
The general strategy of all contemporary studies of primordial magnetic fields (PMF) concerns a search for a self-consistent evolution of the three main characteristics of magnetic fields: the correlation length $\Lambda(\eta)$, the field strength $B(\eta)$ (or the energy density $\rho_B=B^2/2$) and the magnetic helicity density $h=V^{-1}\int d^3x(\mathbf{A\cdot B})$. While a phase transition provides some initial values of these characteristics,
their following evolution leads to issues should be compared with the observations, e.g. of the CMB fluctuations which are sensitive to PMF of the order of a few nG. 

Some new papers have recently appeared on the subject how the intergalactic magnetic fields  originated from the cosmological phase transitions (QCD or EWPT) evolve (see e.g. \cite{Tevzadze:2012kk,Kahniashvili:2012uj}). In the present work we concern a more earlier
epoch before EWPT to estimate some initial parameters for such studies. In particular, we estimated here the chiral anomaly parameter $\Delta \xi=\Delta \mu/T=(\mu_{eR} - \mu_{eL})/T$
arising in hypermagnetic fields just before EWPT which then defines the chiral-magnetic effect
for Maxwellian magnetic field evolution \cite{Boyarsky:2011uy,Tashiro:2012mf}.

We studied the the self-consistent evolution of the hypermagnetic helicity density $\tilde{h}_Y(\eta)$ and the lepton asymmetries $\xi_{eR}(\eta)=\mu_{eR}/T$, $\xi_{eL}(\eta)=\mu_{eL}/T$ in the symmetric phase before EWPT, $\eta < \eta_{EW}$, for the different initial right electron asymmetry $\xi_{eR}(\eta_0)$ (=$10^{-10}$, or $10^{-4}$) that is a {\it free} parameter in the chosen leptogenesis scenario. The left lepton asymmetry $\xi_{eL}\equiv \xi_{\nu_{eL}}$ was fixed at the initial time $\eta_0=M_0/T_{RL}=7\times 10^{13}$, $\xi_{eL}(\eta_0)=0$, when Higgs decays enter equilibrium with the universe expansion at $T_{RL}\simeq 10~TeV$ and involve following left lepton number evolution and a sphaleron influence its value hence the BAU too. We find such influence should be negligible because starting from zero at $\eta_0$ the left lepton number $L_{eL}=(n_{eL} - n_{\bar{e}_L})/n_{\gamma}\sim \xi_{eL}$ has not time to grow before EWPT
remaining small, $\xi_{eL}\ll \xi_{eR}$ (compare Figs.\ref{asymmetryright}, \ref{asymmetryleft} ). Such behaviour of asymmetries was expected a long time ago in \cite{Campbell:1992jd} and stimulated then the choice of a leptogenesis scenario with a non-zero primeval right electron asymmetry as a source of BAU generation in hypermagnetic fields \cite{Giovannini:1997eg,Semikoz:2009ye}. The Abelian anomaly arising in such fields leads to the lepton number violation and evolution of asymmetries $\xi_{eR}$, $\xi_{eL}$. 

The saturation asymmetry level $\Xi_e^{(satur)}=\xi_{eR} + \xi_{eL}/2\approx \xi_{eR}$ turns out to be independent of the initial value $\xi_{eR}(\eta_0)$ in the case of monochromatic helicity spectrum. For the more realistic continuous 
helicity spectrum such saturation level decreases about two orders of magnitude for comparable $\tilde{k}_0\sim \tilde{k}_{max}$, compare Fig.\ref{asymmetryright} and Figs.\ref{asymmetryright3}. This happens again due to the inverse cascade and an accumulation of the negative diffusion influence the growth of $\xi_{eR}$ running modes $0<\tilde{k}<\tilde{k}_{max}$ in integrals $I_{n_s +2}$, $I_{n_s +1}$. It would be interesting in future to check whether the observable BAU $B\sim 10^{-10}$ can be produced in given scenario with hypermagnetic fields accounting for their continuous helicity density spectrum instead of the monocromatic one. 

There is a similarity between equations for the magnetic helicity 
evolution in paper \cite{Boyarsky:2011uy} and the hypermagnetic helicity one in our work and simultaneously
there is a crucial difference of some issues. E.g. Eq. (13) in \cite{Boyarsky:2011uy} is similar to
the first equation (\ref{difsum}) in our work if we put $\xi_{eL}\ll \xi_{eR}$ (as it is confirmed 
by our calculations) hence concelling sphaleron contribution. Analogously Eq. (14) in \cite{Boyarsky:2011uy}
is similar to our Eq. (\ref{energyevolution}). Therefore it is not surprising that tracking solution in \cite{Boyarsky:2011uy},
$\Delta \mu\to \Delta\mu_{tr}$, corresponds to the saturation regime in our plots in Fig. 1.
However, for the chirality flip after EWPT the rate grows over time, $\Gamma_f\sim \eta^2$, due to
the elastic electromagnetic (Rutherford) cross-section $\sigma_{em}\sim \alpha^2/E^2$. This leads
to the following decay of lepton asymmetry and a more faster decrease of the magnetic helicity
driven by this asymmetry. While
the chirality flip before EWPT given by the rate $\Gamma=2\Gamma_{RL}/T$ in (2.12) is constant 
for $\eta\ll \eta_{EW}$ (as well as in the work by Joyce and Shaposhnikov \cite{Joyce:1997uy}) and even vanishes
for $\eta\to \eta_{EW}$. Such behaviour weakens the role of chirality flips
for hypermagnetic helicity evolution comparing with it for magnetic field in Higgs phase
while the hypermagnetic diffusion remains as the main damping mechanism. Nevertheless, for 
a large-scale hypermagnetic field the decrease of helicity density due to diffusion is 
unnoticeable (see in Figs. 4,6).

\section*{Acknowledgments}

We acknowledge Maxim Dvornikov and Alexander Rez for comments.

\section*{Appendix:~~Total system of evolution equations for arbitrary helicity\label{general} }

For completeness we give here the more general system of evolution equations for the spectra of the helicity density $\tilde{h}_Y(\tilde{k},\eta)$ and the energy density $\tilde{\rho}_{B_Y}(\tilde{k},\eta)$ obeying the inequality $\tilde{\rho}_{B_Y}(\tilde{k},\eta)\geq \tilde{k}\tilde{h}_Y(\tilde{k},\eta)/2$ \cite{Biskamp}:
\begin{eqnarray}\label{general}
&&\frac{d\tilde{h}_Y(\tilde{k},\eta)}{d\eta}=-\frac{2\tilde{k}^2}{\sigma_c}\tilde{h}_Y(\tilde{k},\eta) + \left(\frac{4\alpha^{'}(\xi_{eR} + \xi_{eL}/2)}{\pi\sigma_c}\right)\tilde{\rho}_{B_Y}(\tilde{k},\eta),\nonumber\\&&
\frac{d\tilde{\rho}_{B_Y}(\tilde{k},\eta)}{d\eta}=-\frac{2\tilde{k}^2}{\sigma_c}\tilde{\rho}_{B_Y}(\tilde{k},\eta)+ \left(\frac{\alpha^{'}(\xi_{eR} + \xi_{eL}/2)}{\pi\sigma_c}\right)\tilde{k}^2 \tilde{h}_Y(\tilde{k},\eta).
\end{eqnarray} 
For the particular case of the maximum helicity $\tilde{h}_Y(\tilde{k},\eta)=2\tilde{\rho}_{B_Y}(\tilde{k},\eta)/\tilde{k}$ the system (\ref{general}) reads as the single equation Eq. (\ref{conformal}). This system is completed by the kinetic equations for asymmetries $\xi_{eR}(\eta)$, $\xi_{eL}(\eta)$ given by Eqs. (\ref{right}), (\ref{left}). It would be interesting in future to follow from Eqs. (\ref{general}) how the initial non-helical field, $\tilde{h}_Y(\tilde{k},\eta_0)=0$, evolves in the presence of a non-zero initial energy spectrum for which $[d\tilde{h}_Y(\tilde{k},\eta)/d\eta]_{\eta=\eta_0}=(4\alpha^{'}\xi_{eR}(\eta_0)/\pi\sigma_c)\tilde{\rho}_{B_Y}(\tilde{k},\eta_0)\neq 0$. Above in Eq.(\ref{conformal}) we assumed a fully helical field from the beginning while such assumption may be not the case ( see e.g., in \cite{Tevzadze:2012kk}, \cite{Kahniashvili:2012uj}).


\begin{thebibliography}{10}

\bibitem{Betal12}
A.~Brandenburg, D.~Sokoloff  and K.~Subramanian,
\textit{Current Status of Turbulent Dynamo Theory. From Large-Scale to Small-Scale Dynamos},
\textit{Sp. Sci. Rev.} \textbf{169}, 2012,  123-157.

\bibitem{Grasso:2000wj}
  D.~Grasso and H.~R.~Rubinstein,
  \textit{Magnetic fields in the early Universe},
  \textit{Phys. Rept.} \textbf{348} (2001) 163
  [astro-ph/0009061].
  

\bibitem{BK}
F. Krause, F., R. Beck, R., \textit{Symmetry and direction of seed 
magnetic fields in galaxies },
 \newblock  Astron. Astrophy., {\bf 335}, 789 (1998).

\bibitem{Neronov:2009gh}
  A.~Neronov and D.~V.~Semikoz,
  \textit{Sensitivity of gamma-ray telescopes for detection of magnetic fields in intergalactic medium},
  \textit{Phys. Rev.} \textbf{D 80} (2009) 123012
  [arXiv:0910.1920].
%%CITATION = 0910.1920;%%

\bibitem{Neronov:1900zz}
  A.~Neronov and I.~Vovk,
  \textit{Evidence for strong extragalactic magnetic fields from Fermi observations of TeV blazars},
  \textit{Science} \textbf{328} (2010) 73
  [arXiv:1006.3504].

\bibitem{Semikoz:2011tm}
  V.~B.~Semikoz and J.~W.~F.~Valle,
  \textit{Chern-Simons anomaly as polarization effect},
  \textit{JCAP} \textbf{11} (2011) 048
  [arXiv:1104.3106].

\bibitem{Giovannini:1997eg}
  M.~Giovannini and M.~E.~Shaposhnikov,
  \textit{Primordial hypermagnetic fields and triangle anomaly}
  Phys. Rev. \textbf{D 57} (1998) 2186
  [hep-ph/9710234].
 
 
\bibitem{Redlich:1984md}
  A.~N.~Redlich and L.~C.~R.~Wijewardhana,
  \textit{Induced Chern-Simons
  terms at high temperatures and finite densities},
  \textit{Phys. Rev. Lett.} \textbf{54} (1985) 970.
   
\bibitem{Akhmet'ev:2010ba}
  P.~M.~Akhmet'ev, V.~B.~ Semikoz and D.~D.~Sokoloff,
  \textit{Flow of hypermagnetic helicity in the embryo of a new phase in the electroweak transition},
  \textit{JETP Letters} \textbf{91} (2010) 215
  [arXiv:1002.4969].


\bibitem{Semikoz:2012ka}
  V.~B.~Semikoz, D.~ Sokoloff and J.~W.~F.~Valle,
  \textit{Lepton asymmetries and primordial hypermagnetic helicity evolution},
  \textit{JCAP} \textbf{06} (2012) 008 
  [arXiv:1205.3607].

\bibitem{Boyarsky:2011uy}
  A.~Boyarsky, J.~Fr\"{o}hlich and O.~Ruchayskiy,
  \textit{Self-consistent evolution of magnetic fields and chiral asymmetry in the early Universe},
  \textit{Phys. Rev. Lett.} \textbf {108} (2012) 031301 [arXive:1109.3350~[astro-ph]].
  
\bibitem{Boyarsky:2012ex}
A.~Boyarsky, O.~ Ruchayskiy and M. Shaposhnikov,
\textit{Long-range magnetic fields in the ground state of the Standard Model plasma},
\textit{Phys. Rev. Lett.} \textbf{109} (2012) 111602
[arXiv:1204.3604 [hep-ph]] 

\bibitem{Biskamp}
D. Biskamp, \textit{Magnetohydrodynamic Turbulence}, Cambridge University Press, Cambridge, 2003.
 
\bibitem{Dvornikov:2011ey}
  M.~Dvornikov and V.~B.~Semikoz,
  \textit{Leptogenesis via hypermagnetic fields and baryon asymmetry},
  \textit{JCAP} \textbf{02} (2012) 040
  [arXiv:1111.6876];
  \textit{Erratum:} \textit{JCAP} \textbf{08} (2012) E01.

\bibitem{Dvornikov:2012rk}
M.~Dvornikov and V.~B.~Semikoz, 
\textit{Lepton asymmetry growth in the symmetric phase of an
electroweak plasma with hypermagnetic fields versus its washing out by sphalerons}
Phys. Rev \textbf{D87} (2013) 025023.

\bibitem{Zee}
  A.~Zee,
  \textit{Quantum field theory in a nutshell},
  Princeton University Press, Princeton U.S.A. (2010), pg.~270.

\bibitem{Semikoz:2004rr}
  V.~B.~ Semikoz and D.~D.~Sokoloff,
  \textit{Magnetic helicity and cosmological magnetic field},
  \textit{Astron. Astrophys.} \textbf{433} (2005) L53
  [astro-ph/0411496].
  

\bibitem{Semikoz:2003qt}
  V.~B. Semikoz and D.~D. Sokoloff,
  \textit{Large -scale magnetic field generation by alpha-effect driven by collective neutrino-plasma interaction},
  \textit{Phys. Rev. Lett.} \textbf{92} (2004) 131301
  [astro-ph/0312567].
%%CITATION = ASTRO-PH/0312567 1204.3604].;%%


\bibitem{Rubakov}
D.~S.~Gorbunov and V.~A.~Rubakov,
  \textit{Introduction to the theory of the early Universe: Hot Big Bang theory},
  World Scientific Publishing Company, Singapore (2011), pg.~251.

\bibitem{Campbell:1992jd}
  B.~A.~Campbell, S.~Davidson, J.~Ellis and K.~A.~Olive,
  \textit{On the
  baryon, lepton-flavor and right-handed electron asymmetries of the
  universe}
  Phys. Lett. B \textbf{297} (1992) 118
  [hep-ph/9302221].


\bibitem{1983flma....3.....Z}
I.~B. {Zeldovich}, A.~A. {Ruzmaikin} and D.~D. {Sokolov},
\newblock {\em {Magnetic fields in astrophysics}} (New York, Gordon and Breach
  Science Publishers), 1983, 381 p.


\bibitem{Semikoz:2007ti}
  V.~B.~Semikoz and J.~W.~F.~Valle,
  \textit{Lepton asymmetries and the growth of cosmological seed magnetic fields},
  \textit{JHEP} \textbf{03} (2008) 067
  [arXiv:0704.3978].
 
\bibitem{Semikoz:2009ye}
  V.~B.~Semikoz, D.~D.~Sokoloff and J.~W.~F.~Valle,
  \textit{Is the baryon asymmetry of the Universe related to galactic magnetic fields?},
  \textit{Phys. Rev.} \textbf{D 80} (2009) 083510
  [arXiv:0905.3365].
   

\bibitem{Shaposhnikov:2008pf}
M. Shaposhnikov, \textit{ The $\nu MSM$ leptonic asymmetries and properties of singlet fermions},
JHEP \textbf{08} (2008) 008 [hep-ph/0804.4542].

\bibitem{Tashiro:2012mf}
Hiroyuki Tashiro, Tanmay Vachaspati and Alexander Vilenkin,
\textit{Chiral Effects and Cosmic Magnetic Fields},
\textit{Phys. Rev.} \textbf{D 86} (2012) 105033
[arXiv:1206.5549 [astro-ph.CO]].

\bibitem{Joyce:1997uy}
 M. Joyce and M. Shaposhnikov, 
\textit{Primordial magnetic fields, right-handed electrons, and the Abelian anomaly},
\textit{Phys. Rev. Lett.} \textbf{79} (1997) 1193 [astro-ph/9703005]. 
  
\bibitem{Tevzadze:2012kk}
A. Tevladze, L. Kisslinger, A. Brandenburg and T. Kahniashvili,
\textit{Magnetic fields from QCD phase transitions},
Astrophys.J. \textbf{759} (2012) 54 [arXive: 1207.0751 (2012)].

\bibitem{Kahniashvili:2012uj}
T. Kahniashvili, A. Tevladze, A. Brandenburg and A. Neronov,
\textit{Evolution of Primordial Magnetic Fields from Phase Transitions},
\textit{Phys. Rev.} \textbf{D 87} (2012) 083007 [arXive:1212.0596 [astro-ph.CO]]. 

\end{thebibliography}
\end{document}